\documentclass[article,twocolumn,aps,superscriptaddress,longbibliography]{revtex4-1}

\usepackage{graphicx}
\usepackage{dcolumn}% Align table columns on decimal point
\usepackage{bm}% bold math
\usepackage{epstopdf}
\usepackage{float}
\usepackage{amsmath}
\usepackage{xcolor}
\usepackage{comment}
\usepackage[normalem]{ulem}
\graphicspath{{./images/}}
\usepackage{pgfplots}

\usepackage{pgfplots}\pgfplotsset{compat=1.8}
\usepackage{lineno}

\usepackage{tikz}
\usetikzlibrary{calc}

\usepackage{xcolor}
\definecolor{lightblue}{rgb}{0.2,0.2, 0.9}

\newcommand\drawparticle[5]{
    \shadedraw[ball color=#1, draw=none] (axis cs:#2,#3,#4) circle (#5 pt); 
}

\def \radiusparticle{5.5}
\def \zparticle{-0.5}
\def \shiftdoubleoccupation{pi*0.095}

\begin{document}
\title{{Collective excitations of a Bose-Einstein condensate of hard-core bosons and their mediated interactions: from two-body bound states to mediated superfluidity }}
\author{Moroni Santiago-Garc\'ia}
\affiliation{Instituto Nacional de Astrof\'isica, \'Optica y Electr\'onica,
Calle Luis Enrique Erro No.1 Santa Mar\'ia Tonantzintla, Puebla CP 72840, Mexico}
\author{Arturo Camacho-Guardian}
\email{acamacho@fisica.unam.mx}
\affiliation{Instituto de F\'isica, Universidad Nacional Aut\'onoma de M\'exico, Apartado Postal 20-364, Ciudad de M\'exico C.P. 01000, Mexico}

%\linenumbers
\date{\today}
\begin{abstract}
{The exchange of collective modes has been demonstrated to be a powerful tool for inducing superconductivity and superfluidity in various condensed matter and atomic systems. In this article, we study the mediated interactions of collective excitations in an ultracold gas of hard-core bosons.} We show that the induced interaction supports two-body states with energies, symmetries, and a number of bound states strongly dependent on the properties of the hard-core boson gas. The ability to control the nature of the two-body bound states motivates the study of superfluid phases, which we address within the BKT theory. We demonstrate how the superfluid parameters and critical temperatures can be tuned in our system. Our findings may pave the way for future theoretical and experimental studies with ultracold gases and solid-state systems.

\end{abstract}

\maketitle

\section{Introduction}

Induced interactions play a fundamental role in many processes in physics, from phonon-induced Cooper pairing between electrons responsible for superconductivity~\cite{Cooper1956,Bardeen1957} to fundamental interactions in high-energy physics~\cite{yukawa1935interaction,yukawa1937interaction,ruderman1954indirect}. Engineering different mechanisms for tunable induced interactions has received notable attention in atomic and condensed matter physics. In the context of ultracold gases, the study of interactions mediated by the collective excitations of Bose gases has been extensively explored. Specifically, studies of phonon-like excitations of a weakly interacting Bose-Einstein condensate (BEC) have predicted intriguing few- and many-body phases, such as conventional superfluidity~\cite{Wang2005,wang2006strong,pasek2019induced,caracanhas2017fermi,wang2019superfluid,matsyshyn2018p}, topological superfluidity~\cite{Wu2016,Kinnunen2018,Midtgaard2017,Wu2020,zhang2021stabilizing}, bipolarons~\cite{Dehkharghani2018,camacho2018bipolarons,petkovic2022,charalambous2019,ardila2020strong,theel2023crossover,jager2022effect,will2021polaron}, and few-body states~\cite{Naidon2018}. These studies have encouraged further investigation into the intrinsic nature of mediated interactions~\cite{fujii2022,reichert2019} and the interaction between charged particles~\cite{ding2022,astrakharchik2023many}. {However, generating, probing, and measuring experimentally induced interactions in quantum gases remains a complex challenge~\cite{DeSalvo2019,baroni2023mediated}. }

Quantum microscopy delivered a powerful tool to explore phases of matter in optical lattices with site-by-site imaging~\cite{Bakr2009}, motivating the study of Bose lattice polarons~\cite{Colussi2022}, their mediated interactions~\cite{Ding2022a} and strongly correlated physics with lattice Fermi polarons~\cite{Nielsen2021,Kraats2022,Nielsen2022,Grusdt2018,Grusdt2019,Bohrdt2021,Koepsell2021,Pimenov2021,CamachoP}. {The increasing interest on strongly correlated quantum gases in optical lattices is therefore motivated by a parallel development of the abilities to measure such phases with unprecedented control.}

Ultracold gases are useful systems for emulating complex solid-state and condensed matter physics. In particular, when ultracold bosons are tightly confined to a two-dimensional lattice and subject to strong interactions, they exhibit collective excitations that resemble those of spin  systems\cite{bernardet2002analytical}. {In this sense, ultracold gases are referred to as \textit{quantum analogues}, where the system does not correspond to real magnetic materials with true spin excitations. However, the Hamiltonian closely mimics, under certain regimes, the Hamiltonian of their solid-state counterpart, allowing for quantum simulation. }
\begin{figure}[h]
\centering
\includegraphics[width=\columnwidth]{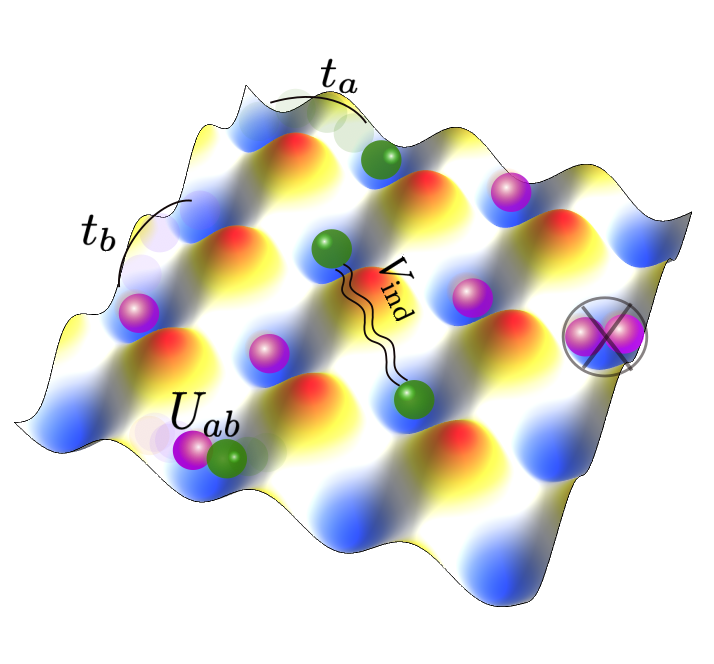}
\caption{A two-dimensional optical lattice confining a gas of hard-core bosons (purple balls) and a second species of atoms (greens balls). The collective excitations of the gas of hard-core bosons mediate an interaction between the minority atoms. }
\label{Fig0}
\end{figure}

{{\it Analog} spin collective excitations can become relevant for a deeper understanding of solid-state magnetic materials.  }In magnetic materials coupled to a conductor~\cite{brataas2020spin}, the interaction between the spins and the electrons in the conductor can induce an effective attraction between the electrons, leading to superconductivity~\cite{Rohling2018,erlandsen2020schwinger,johansen2019magnon,erlandsen2019enhancement}. Similar to conventional superconductivity, which is mediated by phonons (the elementary excitations of the crystal), the collective excitations of the spins, termed magnons, lead to an attractive interaction between electrons. Magnon-mediated superconductivity has garnered considerable attention, particularly for its potential to produce topological 
superconductivity phases~\cite{erlandsen2020magnon,hugdal2020possible}.   

Recent experiments with moir\'e systems in van der Waals heterostructures have opened up new possibilities for studying collective modes analogous to those found in magnetic systems~\cite{Remez2022,JulkuA}. Despite this progress, however, the study of {the collective excitations of hard-core bosons}  and their mediated interactions in ultracold gases remains largely unexplored.

In this article, we study the interaction induced by hard-core bosons and the emergence of two-body bound states, as well as mediated superfluidity. We show that the collective excitations of hard-core bosons, which resemble spin-like excitations, mediate an interaction of a different character than the one induced by weakly interacting BECs. Here, we provide a comprehensive and self-contained study of the mediated interactions by hard-core bosons. 

The remainder of the manuscript is organized in four main parts. In Sec.~\eqref{Model} we detail the model. Following Ref.~\cite{bernardet2002analytical}, we discuss the elementary excitations of the gas of hard-core boson. Then, we derive the Fr\"ohlich like Hamiltonian to describe the coupling between the collective excitations of the gas and a minority species coupled to the gas. This provides the necessary building blocks to derive the mediated interaction due to the exchange of spin-wave like excitations in the medium in Sec.~\eqref{Mediated}. In Sec.~\eqref{Bound}, we study the emergence of two-body bound states and demonstrate that the energy, number, and symmetries of the bound states can be varied over a wide range of parameters. Finally, in Sec.~\eqref{BCS}, we study mediated superfluidity, determining the superfluid parameter, the superfluid fraction, and the critical temperature within the Bardeen-Cooper-Schrieffer theory (BCS) and Berezinskii–Kosterlitz–Thouless (BKT) formalism. {In Sec.~\ref{Experiment}, we discuss the experimental relevance of our study in different contexts. Finally,} we conclude and provide an overlook of our findings in Sec.~\eqref{Conclusions}.

{Our study focuses on quantum gases. However, it is strongly motivated not only by the state-of-the-art experiments in quantum gases, where strongly correlated physics has already been realized in optical lattices, but also by the growing interest and capabilities to design Bose-Fermi Hubbard models in van der Waals heterostructures. These designs exhibit a degree of control that is starting to challenge ultracold gases. Furthermore, recent proposals in magnetic materials to induce topological superconductivity via spin fluctuations have emerged as an alternative avenue for quantum simulation. As a result, our work may serve as a valuable benchmark for future experiments and theories.  }

\section{Model}
\label{Model}

Our system of study consists of a gas of hard-core bosons (majority atoms) coupled to a few identical particles (minority atoms). The binary mixture is confined to a two-dimensional square optical lattice as illustrated in  Fig.~\eqref{Fig0}. 

To make our manuscript self-contained, in this section we detail the system and present a short derivation of the collective excitations of the bosonic gas. We use this formalism to derive the coupling of the hard-core bosons to the second species of atoms in terms of the collective modes of the former. This section gives the theoretical background to address the study of mediated interaction by the exchange of magnon-like excitations. Readers already familiar with this first part can directly jump into Sec.~\eqref{Frohlich}.

\subsection{Collective Excitations: Hard-Core Bosons}
\label{collective}
We start describing the collective excitations of the hard-core bosons. Our starting point is an ultracold gas of bosons confined to a two-dimensional square optical lattice of $N_s$ sites. The Hamiltonian of the bosons is given by
\begin{gather}
\hat H_B=-t_b\sum_{\langle\mathbf r,\mathbf r'\rangle}(\hat b^\dagger_{\mathbf r}\hat b_{\mathbf r'}+\text{h.c})+\\ \nonumber
+\frac{U_{bb}}{2}\sum_{\mathbf r}\hat b^\dagger_{\mathbf r}\hat b^\dagger_{\mathbf r}\hat b_{\mathbf r}\hat b_{\mathbf r}-\mu_b\sum_{\mathbf r}\hat b^\dagger_{\mathbf r}\hat b_{\mathbf r},
\end{gather}
here $\hat b^\dagger_{\mathbf r}$ creates a boson in site $\mathbf r$,  $t_b$ is the nearest neighbour tunneling coefficient, the interaction between the bosons $U_{bb}$ is assumed to be local, and $\mu_b$ is the chemical potential. 

We work in the regime where $U_{bb}\gg t_b$ and assume that double occupation is strictly forbidden. That is, we impose the hard-core constraint  for the majority bosons. Now, following Ref.~\cite{bernardet2002analytical} we introduce spin operators
\begin{gather}
\hat b^\dagger_{\mathbf r}\rightarrow \hat S^\dagger_{\mathbf r},\ \text{and}\,   \hat b_{\mathbf r}\rightarrow \hat S_{\mathbf r}, 
\end{gather}
and map the Hamiltonian to a Heisenberg Hamiltonian of the form
\begin{gather}
 \hat H_B=-t_b\sum_{\langle \mathbf r,\mathbf r'\rangle}(\hat S^\dagger_{\mathbf r}\hat S_{\mathbf r'}+\text{h.c})-\mu_b\sum_{\mathbf r}\hat S_{\mathbf r}^z-\frac{\mu_b}{2} N_s,  
\end{gather}
which no longer explicitly depends on $U_{bb}.$

The mean-field solution is obtained assuming that all sites fill equally, and thus an ansatz for the ground-state is proposed as following
\begin{gather}
 |\Psi\rangle=\prod_{\mathbf r}\left[\sin\left(\frac{\theta}{2}\right)+\cos\left(\frac{\theta}{2}\right)\hat S^\dagger_{\mathbf r}\right]|0\rangle,   
\end{gather}
here the angle $\theta$ is obtained by means of the variational principle $\delta\langle \Psi| \hat H_B|\Psi\rangle/\delta \theta=0.$ Upon minimizing the energy, one obtains for the variational parameter in terms of the chemical potential $\cos\theta=\mu_b/4t_b,$ which gives a filling factor of $n_B=(\cos\theta+1)/2$ and a condensate fraction $n_0=\sin^2\theta/4.$ 

Now, we obtain the collective excitations. This is a two-step procedure where we first rotate our system to align the $z$-axis to the mean-field solution 
\begin{gather}
\begin{bmatrix}
 \hat S^x_{\mathbf r}\\
 \hat S^y_{\mathbf r}\\
 \hat S^z_{\mathbf r}
\end{bmatrix}=
\begin{bmatrix}
\cos\theta & 0 & \sin\theta \\
0 & 1 & 0\\
-\sin\theta& 0 & \cos\theta
\end{bmatrix}
\begin{bmatrix}
 \hat L^x_{\mathbf r}\\
 \hat L^y_{\mathbf r}\\
 \hat L^z_{\mathbf r}
\end{bmatrix}.
\end{gather}
Then we introduce the Holstein-Primakoff transformation by means of the operators $\hat d^\dagger_{\mathbf r}$ and $\hat d_{\mathbf r}$ assumed to be bosonic and given by
\begin{gather}
\hat L^x_{\mathbf r}=\frac{1}{2}(\hat d^\dagger_{\mathbf r}+\hat d_{\mathbf r}), \\ \nonumber
\hat L^y_{\mathbf r}=\frac{1}{2i}(\hat d^\dagger_{\mathbf r}-\hat d_{\mathbf r}), \\ \nonumber
\hat L^z_{\mathbf r}=\frac{1}{2}-\hat d^\dagger_{\mathbf r}\hat d_{\mathbf r}.
\end{gather}
Grouping terms, we write the Hamiltonian as
\begin{gather}
\hat H_B\approx \sum_{\mathbf k} A_{\mathbf k}\left(\hat d^\dagger_{\mathbf k}\hat d_{\mathbf k}+\hat d^\dagger_{-\mathbf k}\hat d_{-\mathbf k}\right)+B_{\mathbf k}\left(\hat d^\dagger_{\mathbf k}\hat d^\dagger_{-\mathbf k}+\hat d_{-\mathbf k}\hat d_{\mathbf k}\right),     
\end{gather}
up to a constant energy shift~\cite{bernardet2002analytical}. Here, the $A_{\mathbf k}$ and $B_{\mathbf k}$ coefficients are given by,
\begin{gather}
A_{\mathbf k}=\frac{1}{2}\left[\frac{\epsilon_{\mathbf k}}{2}(\cos^2\theta+1)+4t_b\right],\\ \nonumber
B_{\mathbf k}=-\frac{1}{4}\sin^2\theta\epsilon_{\mathbf k},
\end{gather}
with $\epsilon_{\mathbf k}=-2t_b(\cos(k_xa)+\cos(k_ya))$ where $a$ is the lattice constant. Due to the quadratic form of the Hamiltonian, we perform a Bogoliubov transformation
\begin{gather}
\hat d_{\mathbf k}=u_{\mathbf k}\hat\gamma_{\mathbf k}-v_{\mathbf k}\hat\gamma^\dagger_{-\mathbf k}\\ \nonumber
\hat d_{-\mathbf k}^\dagger=u_{\mathbf k}\hat\gamma_{-\mathbf k}^\dagger-v_{\mathbf k}\hat \gamma_{\mathbf k}
\end{gather}
that takes the Hamiltonian to its diagonal form, and gives the collective excitations of the hard-core Bose gas $$\hat H_B=\sum_{\mathbf k}\omega(\mathbf k)\hat \gamma_{\mathbf k}^\dagger\hat \gamma_{\mathbf k},$$ where the dispersion of the collective excitations is given by $\omega(\mathbf k)=2\sqrt{A^2_{\mathbf k}-B^2_{\mathbf k}}.$ For small $\mathbf k,$ the dispersion becomes linear
\begin{gather}
\omega(\mathbf k)\approx 2t_ba\sin\theta\sqrt{k_x^2+k_y^2}.
\end{gather}
This dispersion resembles the Bogoliubov dispersion of weakly interacting BECs; however, in this case, the interaction does not longer depend on the boson-boson interaction but solely on the tunneling coefficient and the angle $\theta.$ 

{We remark that although hard-core bosons can be  described by a Heisenberg Hamiltonian with spin-wave excitations, this does not convert the resulting collective excitation into {\it real} magnons  quasiparticles carrying real spin. Due to the close analogy to spin-excitations (magnons), the collective excitations of hard-core bosons are, however, commonly coined spin-waves or magnons, inheriting the terminology of magnetic materials. }

\subsection{ Fr\"ohlich Hamiltonian}
\label{Frohlich}
A second species of atoms (minority) is  placed in the two-dimensional optical lattice and couples to the hard-core bosons. The kinetic energy term of the second species is given by
\begin{gather}
 \hat H_{a}=-t_a\sum_{\langle\mathbf r,\mathbf r'\rangle}\hat a^\dagger_{\mathbf r}\hat a_{\mathbf r'}+\text{h.c.},   
\end{gather}
here $t_a$ is the nearest-neighbour tunneling coefficient and $\hat a^\dagger_{\mathbf r}$ creates an $a$ atom  in a given site $\mathbf r$.  The coupling between the $a$ and $b$ atoms is short-ranged and local, described by the Hamiltonian
\begin{gather}
\hat H_{ab}=U_{ab}\sum_{\mathbf r} \hat n_{b}(\mathbf r)\hat n_a(\mathbf r),  
\end{gather}
where $\hat n_{a,b}(\mathbf r)$ are the number operator of the $a/b$ atoms respectively. Here, $U_{ab}$ can be tuned by means of a Feshbach resonance~\cite{RevModPhys.82.1225}.

To make further progress, we are required to write the former interaction in terms of the collective excitations of the hard-core bosons. For such purpose, we write the Hamiltonian in terms of the $\hat{d}$ bosonic operators of the Holstein-Primakoff transformation
\begin{gather}
\hat H_{ab}=\frac{U_{ab}}{2}(1+\cos(\theta))\sum_{\mathbf r}\hat n_a(\mathbf r)-\\ \nonumber
-\frac{U_{ab}\sin\theta}{2\sqrt{N_s}}\sum_{\mathbf k,\mathbf q}\left[\hat d^\dagger_{-\mathbf q}+\hat d_{\mathbf q}\right]\hat a^\dagger_{\mathbf k+\mathbf q}\hat a_{\mathbf k}.
\end{gather}
The first term can be intuitively understood as the conventional mean-field correction $U_{ab}n_B\sum_{\mathbf r}\hat n_a(\mathbf r)$, whereas the second line can be written in terms of the collective excitations $\hat \gamma$ as follows
\begin{gather}
\hat H_{F}=\frac{1}{\sqrt{N_s}}V_{F}(\mathbf q)\left[\hat \gamma^\dagger_{-\mathbf q}+\hat \gamma_{\mathbf q}\right]\hat a^\dagger_{\mathbf k+\mathbf q}\hat a_{\mathbf k},
\end{gather}
with the interaction between the $a$ atoms and the collective excitations of the hard-core bosons
\begin{gather}
 V_F(\mathbf q)=-\frac{1}{2}U_{ab}(u_{\mathbf q}-v_{\mathbf q})\sin\theta.   
\end{gather}
For the Fr\"ohlich-like Hamiltonian, we only retain as usual linear terms in the collective excitations. 
\section{Mediated Interaction}
\label{Mediated}
We have now collected the elements to derive the interactions between the $a$ atoms mediated by the collective excitations of the hard-core boson gas. Our starting point is the Hamiltonian written in terms of the {$\hat{d}$} operators, after having performed the Holstein-Primakoff transformation.

We derive the mediated interaction following a field theoretical approach where we trace over the collective modes of the hard-core bosons~\cite{altland2010condensed}. We introduce the action
\begin{gather}
 S(\psi_a^*,\psi_a,\phi_d^*,\phi_d)=S_{d}(\phi_d^*,\phi_d)+S_{a}(\psi_a^*,\psi_a) \\ \nonumber 
 +S_{ad}(\psi_a^*,\psi_a,\phi_d^*,\phi_d).
\end{gather}
here, the action  $S_{d}(\phi_d^*,\phi_d)$ depends on the bosonic fields $\phi_d$ and is given by
\begin{gather}
 S_d(\phi_d^*,\phi_d)=\sum_{k}\Phi_d^*(k)\cdot[-\mathcal G^{-1}_d(k)]\cdot\Phi_d(k),
 \end{gather}
 with $\Phi_d^*(k)=[\phi_d^*(k),\phi_d(-k)]$. The Green's function of the d-bosons is given by
\begin{gather}
-\mathcal G^{-1}_d(k)=\begin{bmatrix}-i\nu_n+A_{\mathbf k} & B_{\mathbf k} \\ B_{\mathbf k} & i\nu_n+A_{\mathbf k}\end{bmatrix}
\end{gather}
where $k=(\mathbf k,i\nu_n)$ with $\nu_n$ a bosonic Matsubara frequency.

The action accounting for the coupling between the $a$ atoms and the bosons is given by
\begin{gather}
S_{ab}(\psi_a^*,\psi_a,\phi_d^*,\phi_d)=\frac{1}{\sqrt{N_s}}\sum_{q}\Phi_d^*(q)\cdot J_a(q)+J_a^*(q)\cdot \Phi_d(q),    
\end{gather}
{with the vector $J_a^*(q)=(\frac{U_{ab}}{2}\sum_k\psi_a^*(k)\psi_a(k+q),\frac{U_{ab}}{2}\sum_k\psi_a^*(k)\psi_a(k+q)).$} Finally, the action of the $a$ atoms is 
$$S_a(\psi^*,\psi)=-\sum_{q}\psi_a(q)(i\nu_n-\epsilon_{\mathbf q})\psi_a(q).$$ 

 From the grand partition function $Z=\int \text{d}\phi_d\text{d}\psi_a\text{d}\phi_d^*\text{d}\psi_a^*\exp[- S(\psi_a^*,\psi_a,\phi_d^*,\phi_d)],$ we perform the Gaussian integral over the d-fields which traces out these degrees of freedom. The effective action depending only on the a-field is $S_{\text{eff}}(\psi_a^*,\psi_a)=S_{a}(\psi_a^*,\psi_a)+S_{\text{ind}}(\psi_a^*,\psi_a),$ where $S_{\text{ind}}(\psi_a^*,\psi_a),$ arises from the trace and is given by
\begin{gather}
S_{\text{ind}}(\psi^*,\psi)= J_a^*(q)\cdot \mathcal G_d(q)\cdot J_a(q)=\\ \nonumber=
\frac{U_{ab}^2}{2N_s}\sum_{q}\rho_a(q)\frac{A_{\mathbf q}-B_{\mathbf q}}{i\nu_n^2-(A^2_{\mathbf q}-B_{\mathbf q}^2)}\sin^2\theta\rho_a(q),
\end{gather}
here $\rho_a(q)=\sum_{k}\psi_a^*(k)\psi_a(k+q).$  This term is quartic in the $a$ fields and corresponds to a mediated interaction due to the exchange of a collective mode of the hard-core Bose gas.

To understand the character of the mediated interaction let us focus on the static interaction, given by
\begin{gather}
\label{Vind}
V_{\text{ind}}(\mathbf q)=-\frac{U_{ab}^2}{4t_b+\cos^2\theta\epsilon_{\mathbf q}}\sin^2\theta,
\end{gather}
as expected, the induced interaction is quadratic in $U_{ab}$  and linear in the condensate density $n_0\propto \sin^2\theta$. Interestingly, the zero-momentum limit of the induced interaction is independent of $\theta,$ that is, it does not depend on the filling factor of the bosons,
\begin{gather}
\lim_{\mathbf q\rightarrow{0}}V_{\text{ind}}(\mathbf q)\propto-\frac{U_{ab}^2}{t_b}.    
\end{gather}
This induced interaction exhibits similarities and striking differences with the interaction mediated by weakly interacting BECs~\cite{Heiselberg2000,Camacho-Guardian2018,scazza2022repulsive}. Firstly, as in the case of phonon-mediated interaction, the zero-energy momentum of the induced interaction is independent of the density of the medium. However, for a weakly interacting BEC ($U_{bb}\ll t_b$), the limit is given by $\lim_{\mathbf q\rightarrow{0}}V_{\text{ind}}(\mathbf q)=-\frac{U_{ab}^2}{U_{bb}}$ and is independent of $t_b$. For hard-core bosons, the zero-energy momentum limit is, of course, independent of $U_{bb}$ and scales inversely proportional to $t_b$. Furthermore, we also note that the potential in Eq.~\eqref{Vind} is symmetric around $\theta=\pi/2$, that is, invariant under the exchange of $n_B\rightarrow (1-n_B)$, reflecting the particle-hole symmetry of the underlying Hamiltonian governing the bosonic bath.

\begin{figure}[h]
\centering
\includegraphics[width=\columnwidth]{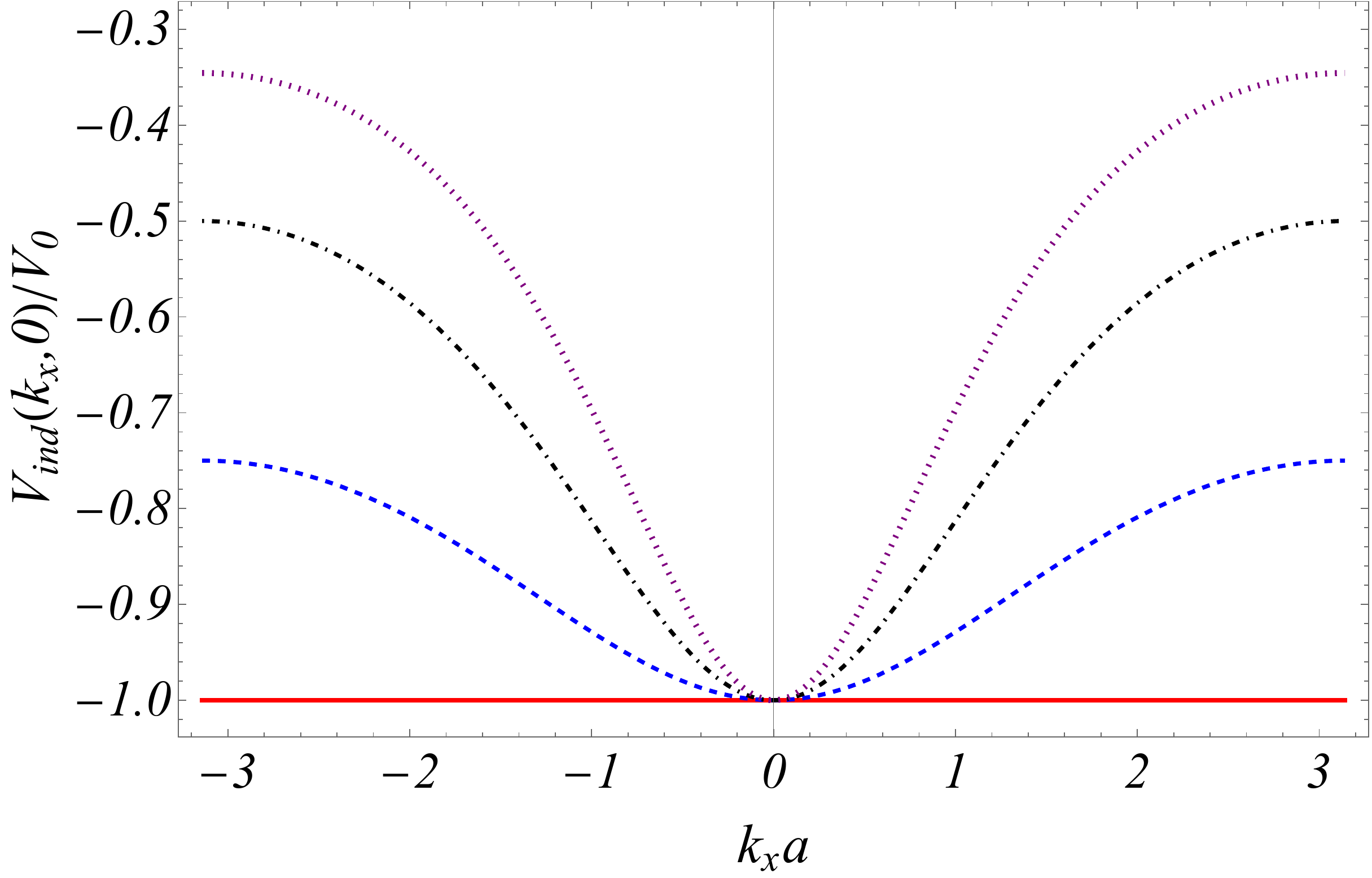}
\caption{Mediated interaction as a function of $k_xa$ for fixed $k_ya=0$. The solid red curve corresponds to half-filling ($\theta=\pi/2$), the blue dashed lines gives the induced interaction for $\theta=\pi/3$, whereas $\theta=\pi/4$ is shown by the dot-dashed black line. Finally, the purple line illustrates the mediated interaction for $\theta=\pi/5$.}
\label{v_ind}
\end{figure}

To gain intuition about the form of the induced interaction, we show in Fig.~\eqref{v_ind} a cross-section of $V_{\text{ind}}(\mathbf q)$ for $k_y=0$ and as a function of $k_x$ for several values of angle $\theta$. The solid red line corresponds to half-filling $(\theta=\pi/2)$. In this case, the induced interaction becomes independent of $\mathbf{k}$ and gives a short-ranged purely on-site interaction. With varying angle, we show that the induced interaction acquires more structure, as shown for $\theta=\pi/3$ (dashed blue), $\theta=\pi/4$ (dot dashed black), and $\theta=\pi/5$ (purple dots). For clarity, we show in units of $V_0=U_{ab}^2/4t_b.$

Figure~\eqref{v_ind} illustrates the strong dependence of the induced interaction on the filling factor of the hard-core boson gas. In particular, the character of the mediated interaction can be controlled via the angle $\theta$. For  $\theta=\pi/2$ the induced interaction is short-ranged and local (contact interaction) while away from half-filling the mediated interaction acquires spatial structure and becomes a finite range-potential where atoms scatter beyond the on-site interaction term. That is, the momentum dependence of the induced interaction in momentum space implies that the interaction in real space is no longer solely on-site interaction.

 {We note that the perfect $s$-wave potential holds only within the static approximation. Retardation effects lead to contributions of higher orders in the partial wave expansion of the potential. However, as long as retardation effects are negligible, the dominant contribution is the $s$-wave

\begin{figure*}[t]
\centering
\includegraphics[width=2\columnwidth]{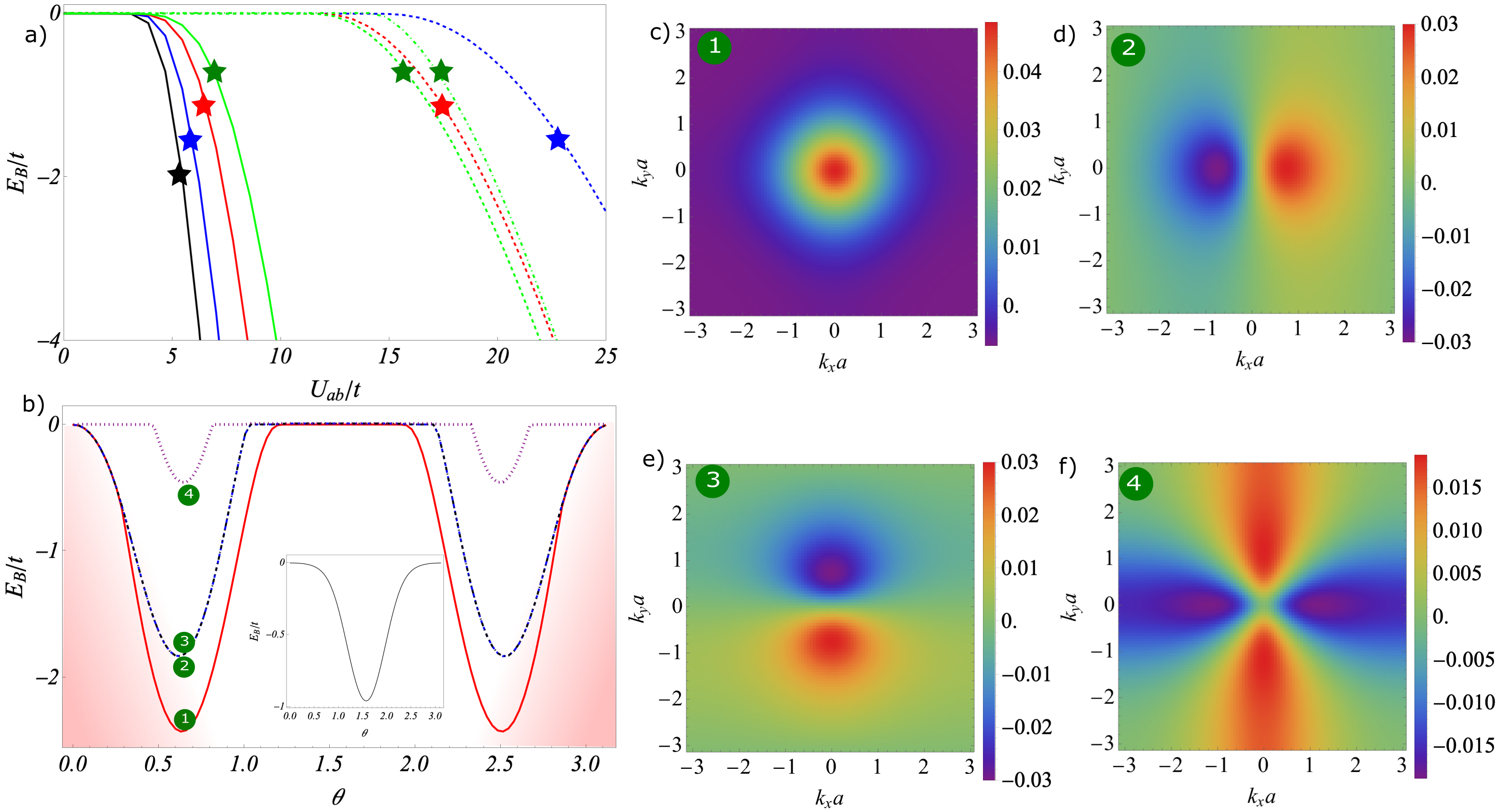}
\caption{Two-body bound states. (a) Binding energies as a function of $U_{ab}/t$ for $\theta=\pi/2$ (black) , $\theta=\pi/3$ (blue), $\theta=\pi/4$ (red) and $\theta=\pi/5$ (green). The solid lines give the lowest bound state for each angle whereas the dashed lines correspond to excited (bound) states. {The stars mark our estimates for the breakdown of the static approximation. } Panel (b) shows the binding energies as a function of $\theta$ for $U_{ab}/t=20$. The red line gives the first excited state. Above the first excited state, we obtain two degenerate states (blue and black on top of each other) and a fourth excited bound state marked by the purple line. The inset gives the lowest bound state energy for $U_{ab}/t=5$. {The red area corresponds to energies beyond the validity of our assumptions in Eq.~\eqref{Estimate}.} Panels (c) to (f) show the real part of the wave function for the points  (1)-(4) marked by the green circles in panel (b) giving different symmetries: from $s-$ to $p-$ and $d-$ wave.  }
\label{Et}
\end{figure*}

Interestingly, the range of the potential is maximal for small filling or close to the unit filling factor. Due to the particle-hole symmetry, the range is symmetric around half-filling and reaches its minimum at half-filling, where the potential becomes local. We attribute this momentum independence to the fact that at half-filling, the minority atom has an equal probability of colliding with both a hole and a particle. In contrast, away from half-filling, there is an imbalance of particles and holes, resulting in a momentum-dependent interaction~\cite{Zoe}. 

}

The induced interaction can be tuned from a short-ranged interaction to a long-ranged interaction by adjusting the filling factor. This contrasts with the case of weakly interacting bosons, whose induced interaction is long-ranged. Moreover, the collective modes are independent of the boson-boson interactions. They  can be tuned simply by varying the hopping parameter, which can be experimentally modified using the confining optical lattice.

{The ability to tune the properties of the confinement allows for adjusting the tunneling coefficients, $t_b$ and $t_a$. We expect that when $t_b \gg t_a$, retardation effects are suppressed because the dynamics of the bosons are much faster than those of the minority atoms. Consequently, the exchange of their collective modes can be considered instantaneous. On the other hand, in the opposite regime where $t_b \ll t_a$, the dynamics of the minority atoms are faster than those of the host bosons, leading to enhanced retardation effects. In our work, we set $t_b = t_a$ and discuss the validity of our approximations.

%In particular, by taking $t_b\neq t_a$, one can enhance or suppress the retardation effects by making the dynamics of the $a$ atoms much faster/slower than the speed of sound of the hard-core bosons. This allows for the exploration of different regimes where retardation may be further suppressed or may play an important role.}
%The ability to tune the range of the induced interaction with spin-wave-like excitations contrasts with the case of weakly interacting bosons, whose induced interaction is more limited.
}

\section{Bound states}
\label{Bound}
Motivated by the possibility of tuning the intrinsic character of the mediated interaction, we turn our attention to the study of the two-body bound states appearing as a consequence of the induced interaction. In this case, we study the Schr\"odinger equation for a pair of $a$-atoms interacting via the static mediated interaction, that is,
\begin{gather}
 E\psi(\mathbf k)=2\epsilon_{\mathbf k}\psi(\mathbf k)+\frac{1}{N_s}\sum_{\mathbf q}V_{\text{ind}}(\mathbf k-\mathbf q)\psi(\mathbf q) 
 \end{gather}
where $\psi(\mathbf k)$ is the wave function of the relative coordinate.  We have written the Schr\"odinger equation in its conventional form in momentum space.

Figure~\eqref{Et} summarises the main features of the two-body bound states as a function of the tuneable parameters of the system. 

First, Fig.~\eqref{Et}(a) shows the binding energy $E_b$ defined as $E=E_b-2\epsilon_{\mathbf k=0}$ as a function of $U_{ab}/t$ for several filling factors. Here, we have taken $t_a=t_b=t.$ For $\theta=\pi/2$ (solid black), we only find one bound state regardless of how large $U_{ab}/t$ is tuned. This is a consequence of the nature of the induced interaction at half-filling: the contact interaction supports a single bound state. On the other hand, for $\theta=\pi/3$ (blue), $\theta=\pi/4$ (red), and $\theta=\pi/5$ (green), we also find a lower-lying bound state with an energy slightly above the bound state for $\theta=\pi/2$. However, away from half-filling, we find that the induced interaction supports multiple bound states. Therefore, giving spatial structure to the mediated interactions yields a rich family of two-body bound states absent at $\theta=\pi/2$.

{Our approach neglects retardation effects, meaning that we assume the mediated interaction to be instantaneous. This results in a frequency-independent mediated interaction. Physically, this approximation can only hold when the typical energy of the collective mode remains larger than the binding energy of the two-body bound state
\begin{gather}
\label{Estimate}
 c>\sqrt{\frac{2|E_B|}{m_{\text{eff}}}},   
\end{gather}
with $c=2ta\sin\theta$ the speed of sound of the BEC of hard core bosons, and $m_\text{eff}\approx 1/ta^2$ the effective mass of the minority atoms. This equation in other words, gives an expression between the typical velocities of the collective mode (speed of sound) and the velocities of the two-body bound state.

In Fig.~\eqref{Et} (a) the stars mark the breakdown of the induced  interaction. Energies below these marks start to become unreliable as retardation effects may become relevant. We note that since the speed of sound of the BEC depends on the filling factors, the position of the stars also has a $\theta$ dependence. This is further illustrated in Fig.~\eqref{Et} (b), where the red area also shows the regimes where the predicted energies lie beyond the validity of our approximations. We should stress that the failure of the instantaneous interaction is not a sharp transition, and the marks and shadowed areas are intended to provide a qualitative guide to the validity of our main assumption.

}

The role of the filling factor in the emergence of many two-body bound states is clarified in Fig.~\eqref{Et}(b). Here, we plot the second to fifth lowest bound states as a function of $\theta$ for $U_{ab}/t=20$. First, as discussed previously, the emergence of a second bound state is very suppressed close to half-filling due to the character of the induced potential. On the other hand, we observe several bound states at different angles, whose energy becomes maximal close to $\pi/5$. In this case, we find a non-degenerate second lowest bound state (solid red line), two degenerate states marked by the blue and black dashed lines (which overlap), and an additional bound state with a smaller binding energy shown in purple. The energy of the lowest bound state is illustrated in Fig.~\eqref{Et}(b, inset) and shows that the lowest bound state energy lies far below the energy of the highest bound states.

The nature of these bound states is unveiled by the wave-function of the states, which reveals the symmetries of the states. In our case, we find that the lowest bound state is always spherically symmetric (not shown). However, the excited states do acquire structure. In Fig.~\eqref{Et}(c)-(f), we show the real part of the wave-function over the first Brillouin zone for the bound states marked by the green circles in Fig.~\eqref{Et}(b), around $\theta=\pi/5$, where the excited bound state peaks. The first excited state (second lowest bound state) is also spherically symmetric. The second and third excited states are degenerate in energy and exhibit a $p$-wave symmetry. Finally, the highest excited bound state possesses a $d$-wave symmetry.

The symmetry of the wavefunction is intrinsically linked to the statistics of the second species of atoms. For bosonic atoms or fermions in a different internal state, the allowed symmetries are the $s$- and $d$-wave wavefunctions. On the other hand, for fermions (spin-polarized), the wavefunction is required to be anti-symmetric in $\mathbf k,$ that is, $\psi(\mathbf k)=-\psi(-\mathbf k),$ the permitted states are therefore the $p-$ wave states in Fig.~\eqref{Et} (d) and (e). 

{Our study shows that the energy, number, and symmetry of the two-body bound states can be tuned over a wide range of parameters. This flexibility is inherited from the range of the induced interaction, which can be varied from short- to long-ranged. This opens the door for studying mediated interactions and their effects, tuning different features than those typically accessible with a weakly interacting BEC.
}

 We note that, although multiple bound states only appear away from half-filling, the most robust two-body state appears at $\theta=\pi/2.$

{The advent of quantum microscopes has provided a powerful tool for resolving atoms loaded site by site in optical lattices~\cite{mazurenko2017cold}. This tool is essential for studying strongly correlated phases in optical lattices~\cite{greif2016site} and has revealed hidden orders in doped Hubbard systems~\cite{Hilker2023}. Furthermore, the ability to probe two-body correlations has allowed for the experimental demonstration of fractional quantum Hall states with ultracold atoms~\cite{leonard2023}. An interesting avenue for future exploration is the investigation of two-body correlations in real space for bound states.
}

In homogeneous and lattice systems, weakly interacting Bose-Einstein condensates have proven to provide an effective mechanism to induce superfluid phases including $s-$ wave and topological superfluidity~\cite{Wang2005,wang2006strong,pasek2019induced,caracanhas2017fermi,wang2019superfluid,matsyshyn2018p,Wu2016,Kinnunen2018,Midtgaard2017,Wu2020,zhang2021stabilizing}. Spin-wave mediated interactions also provide an attractive interaction, which we have demonstrated supports several two-body bound states.

Now, in the following section, we explore the presence of superfluidity mediated by the exchange of collective excitations.

\section{Mediated Superfluidity}
\label{BCS}
Having understood the consequence of the mediated interactions in the emergence of two-body bound states, we now explore the possibility of inducing superfluidity. Specifically, we consider a binary and spin-balanced mixture of fermions embedded in a gas of hard-core bosons. The effective Hamiltonian of the fermions is given by
\begin{gather}
\hat H_{\text{eff}}=-t\sum_\sigma\sum_{\langle \mathbf r,\mathbf r'\rangle}(\hat a^\dagger_{\mathbf r\sigma}\hat a_{\mathbf r'\sigma} +\text{h.c})-\mu\sum_{\mathbf r}\hat a^\dagger_{\mathbf r\sigma}\hat a_{\mathbf r\sigma}+\\ \nonumber
+\sum_{\mathbf r,\mathbf r'}V_{\text{ind}}(\mathbf r-\mathbf r') \hat a^\dagger_{\mathbf r\uparrow}\hat a^\dagger_{\mathbf r'\downarrow}\hat a_{\mathbf r'\downarrow}\hat a_{\mathbf r\uparrow}.  
\end{gather}
here, the operators $\hat a^\dagger_{\mathbf r\sigma}$ create a fermion with spin $\sigma=\uparrow/\downarrow$ in a given site $\mathbf r$. We consider a chemical potential $\mu$ independent of the spin $\sigma$. The fermions only interact via  the interaction  $V_{\text{ind}}(\mathbf r-\mathbf r')$  discussed previously.

For this purpose, we write the Hamiltonian in the standard BCS form
\begin{gather}
\hat H_{\text{BCS}}=\frac{1}{2}\sum_{\mathbf k}[\hat a^\dagger_{\mathbf k\uparrow}\,\hspace{.1cm} \hat a_{-\mathbf k\downarrow}]\cdot\begin{bmatrix}\xi_{\mathbf k} & \Delta_{\mathbf k} \\ \Delta_{\mathbf k}^*& -\xi_{\mathbf k}\end{bmatrix}\cdot \begin{bmatrix}\hat a_{\mathbf k\uparrow} \\ \hat a^\dagger_{-\mathbf k\downarrow}\end{bmatrix} ,  \end{gather}
where the superfluid gap and number equations
\begin{gather}
\label{EqBCS}
\Delta_{\mathbf k}=-\frac{1}{2N_s}\sum_{\mathbf q}V_{\text{ind}}(\mathbf k-\mathbf q)\frac{\Delta_{\mathbf q}}{E_{\mathbf q}}\tanh\left({\frac{E_{\mathbf q}}{2T}}\right), \\
n=\frac{1}{2}\left(1-\frac{1}{N_s}\sum_{\mathbf q}\frac{\xi_{\mathbf q}}{E_{\mathbf q}}\tanh\left({\frac{E_{\mathbf q}}{2k_BT}}\right)\right),
\end{gather}
respectively. Here, the dispersion of the BCS superfluid is given by $E_{\mathbf k}=\sqrt{\xi_{\mathbf k}^2+\Delta_{\mathbf k}^2}.$ We include the Hartree-Fock contribution $\xi_{\mathbf k}=\epsilon_{\mathbf k}+\Sigma_{\mathbf k}-\mu$ with the self-energy term given by
\begin{gather}
\Sigma_{\mathbf k}=\sum_{\mathbf q}\frac{\left[V_{\text{ind}}(\mathbf 0)-V_{\text{ind}}(\mathbf k-\mathbf q)\right]}{2N_s}\left[1-\frac{\xi_{\mathbf q}}{E_{\mathbf q}}\tanh\left({\frac{E_{\mathbf q}}{2k_BT}}\right)\right].    
\end{gather}
Here $k_B$ is the Boltzmann constant, and $T$ the temperature.

In addition, we introduce the superfluid stiffness parameter
\begin{gather}
 J_{i,j}=\frac{1}{4N_sa^2}\sum_{\mathbf k}\left(n(\mathbf k)\frac{\partial^2\epsilon_{\mathbf k}}{\partial k_i\partial k_j}-Y(\mathbf k)\frac{\partial\epsilon_{\mathbf k}}{\partial k_i}\frac{\partial\epsilon_{\mathbf k}}{\partial k_j} \right),   
\end{gather}
where $n(\mathbf k)=|u_{\mathbf k}|^2f(E_{\mathbf k})+|v_{\mathbf k}|^2(1-f(E_{\mathbf k})),$ where $u_{\mathbf k},v_{\mathbf k}$ are the standard BCS Bogoliubov factors $u_{\mathbf k}^2=(1+\xi_{\mathbf q}/E_{\mathbf k})/2$ and $u_{\mathbf k}^2+v_{\mathbf k}^2=1$~\cite{Midtgaard2017}. And the Yoshida function $Y(\mathbf k)=1/4k_BT\text{sech}(E_{\mathbf k}/2T).$

{Guided by our studies of the two-body bound states, we restrict our calculations to the regime where the static approximation remains valid. We can rephrase this condition as:  $c>\sqrt{2\frac{\Delta_0}{m_{\text{eff}}}}$.   
In addition, we work in the weakly interacting regime for the impurity-impurity mediated interaction to ensure that the weak-coupling BCS-theory holds. Here, we take $\left|n_FV_{\text{ind}}(\mathbf 0)/\epsilon_F\right|<1.$ This condition is also important to guarantee that the Fermi gas does not modify the collective excitations of the BEC of hard-core bosons. A strongly interacting theory lies beyond the scope of our manuscript. Here we restrict to value of the relevant parameters to regimes where our assumptions remain valid.

}
{\it Superfluidity.-} We are now in position to study the emergence of superfluid phases. From our previous analysis of the two-body bound states, we expect the $s-$ wave to be the dominant contribution to the superfluid parameter. We find numerically that, indeed, for the parameters explored the gap $\Delta_{\mathbf k}$  is essentially independent of $\mathbf k.$

\begin{figure}[h]
\centering
\includegraphics[width=\columnwidth]{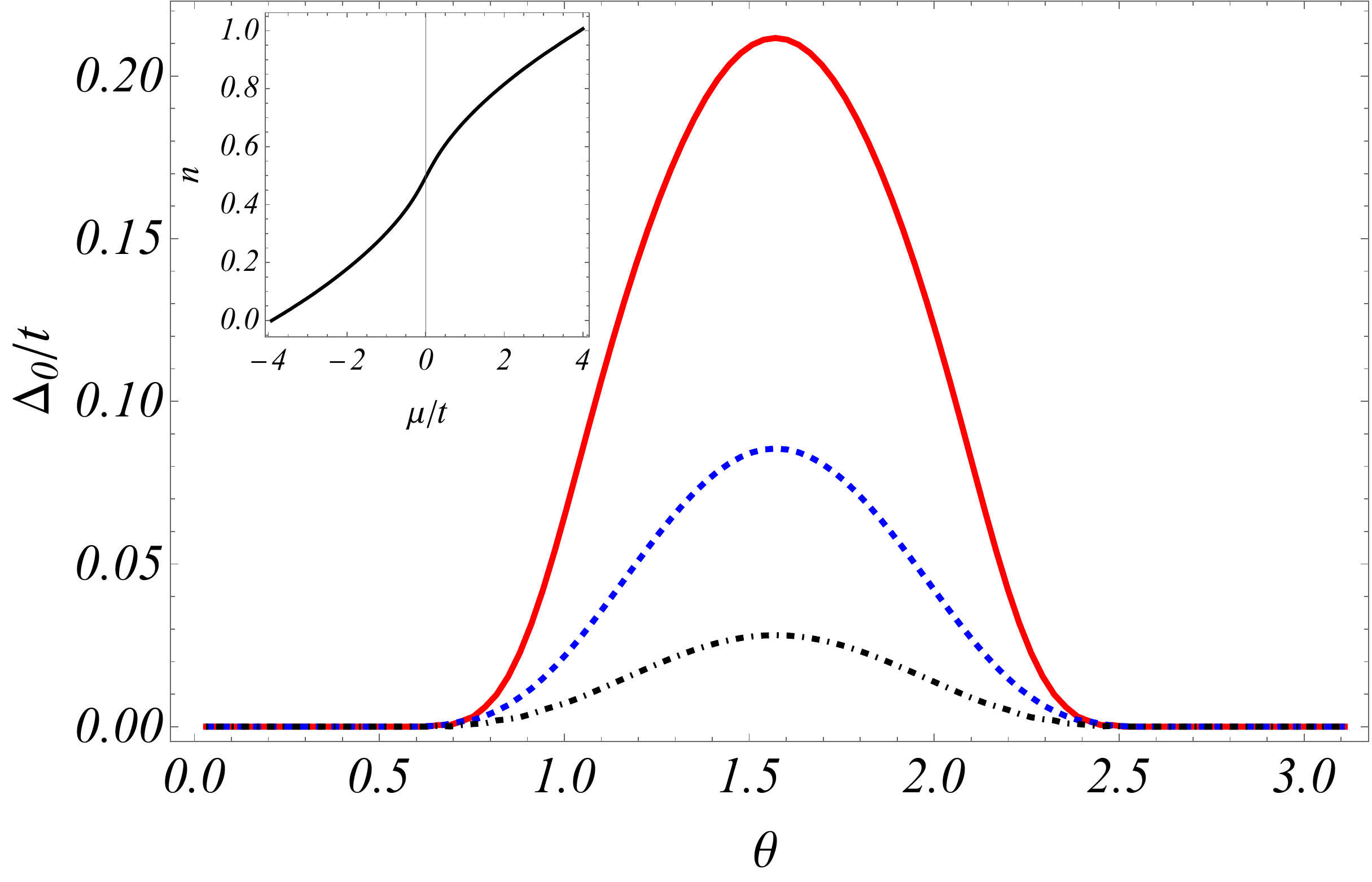}
\caption{Superfluid gap as a function of $\theta$ for $\mu/t=-2$ (dot-dashed black line) and $\mu/t=-1$ (dashed blue line) and $\mu/t=0$ (solid red line). (Inset) Fermions per site as a function of the chemical potential $\mu/t$ for $\theta=\pi/2$. }
\label{Fig1}
\end{figure}

We now explore the emergence of superfluid phases as a function of the angle $\theta$ and at zero temperature. In Fig.~\eqref{Fig1}, we show the superfluid parameter $\Delta_0$ for several values of the fermion chemical potential $\mu$, with $U_{ab}/t$ fixed at 2.5. This value of the fermion-boson scattering interaction lies deep within the range of validity of the static approximation for the two-body problem as shown in Fig.~\ref{Et}(a). We observe the same qualitative behavior of the gap parameter as in the lowest two-body bound state studied previously: the superfluid parameter is maximal at $\pi/2$ and decreases away from half-filling of bosons. {As discussed previously, while the range of the induced interaction is maximal at low filling factors and close to the unit filling factor, the strength of the induced interaction is strongly suppressed due to the small density of particles/holes carrying the mediated interaction. As a result, superfluidity is essentially absent in these limits.} Finally, at large angles, the superfluid gap vanishes. We show results for $\mu/t=0$ (solid red line), $\mu/t=-1$ (dashed blue line), and $\mu/t=-2$ (dot-dashed black line). As an inset, we show the fermion density as a function of the chemical potential $\mu$. We find that the number of fermions $n$, given in terms of $\mu$, is largely independent of $\theta$.

The superfluid parameter primarily arises from $s$-wave scattering and peaks around $\theta=\pi/2$. The dependence on $\theta$ can also be intuitively understood from the condensate fraction of bosons, which scales with the filling factor as $n_0=n_B(1-n_B)=\sin^2\theta/4$. That is, the condensate fraction becomes maximal at $\theta=\pi/2$ and induces an attractive interaction more efficiently around this angle. Away from half-filling of bosons, the induced interaction becomes suppressed and unable to effectively induce fermion $s$-wave pairing.

Now, we will explore the stiffness parameter as a function of temperature. In this case, we keep $U_{ab}/t=2.5$ and the bosonic filling factor $\theta=\pi/2$ fixed. We show the same values of the chemical potential of the fermions as before, using the same color codes: $\mu/t=0$ (solid red line), $\mu/t=-1$ (dashed blue  line), and $\mu/t=-2$ (dot-dashed black line).

\begin{figure}[H]
\centering
\includegraphics[width=\columnwidth]{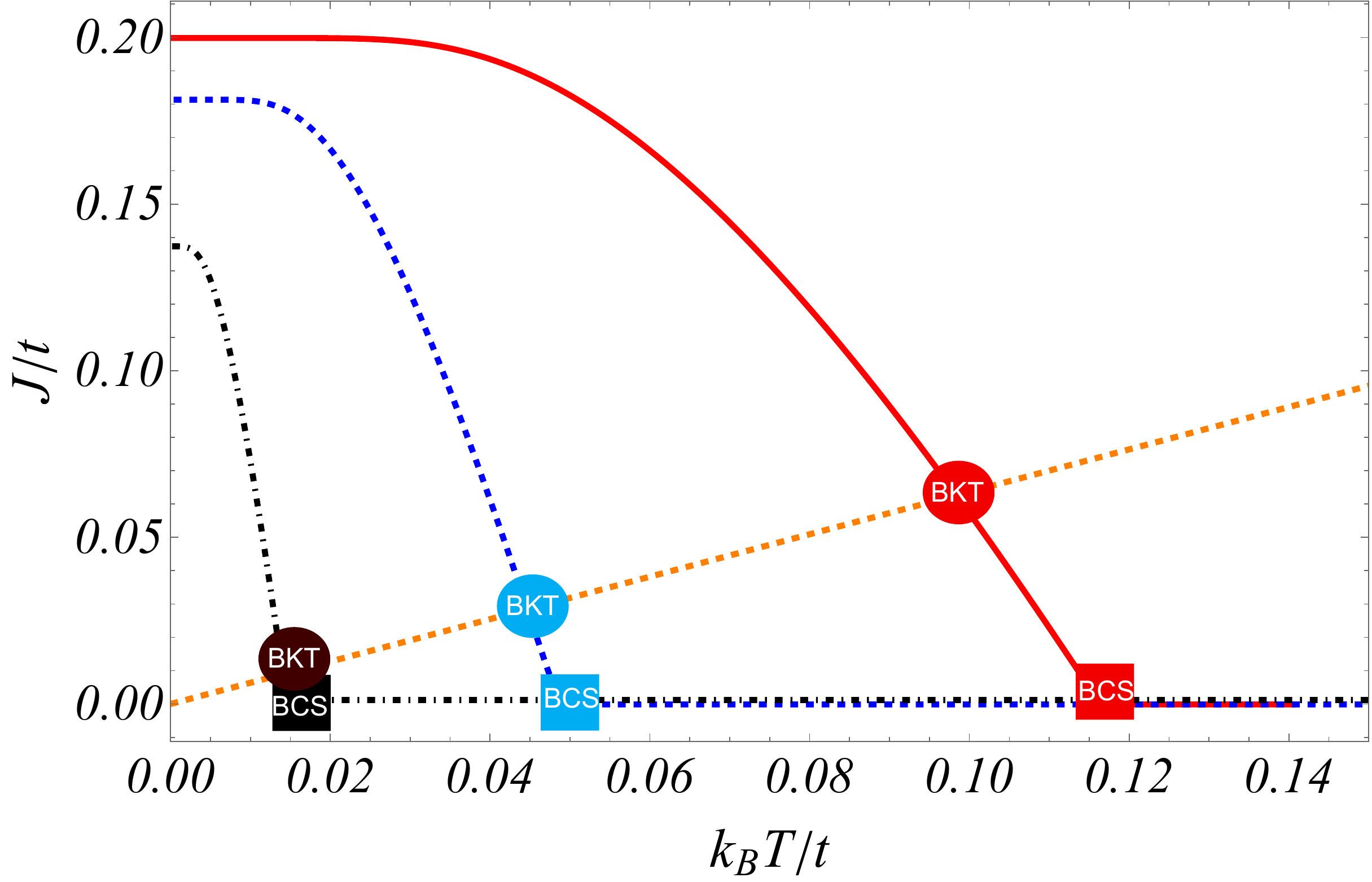}
\caption{The stiffness parameter is shown as a function of temperature for three different values of the chemical potential of the fermions: $\mu/t=-2$ (dot-dashed black line), $\mu/t=-1$ (dashed blue line), and $\mu/t=0$ (solid red line). The critical BCS temperature is marked by filled squares in the figure, as the stiffness parameter vanishes above this temperature. Additionally, the filled circles indicate the critical BKT temperature given by Eq.~\eqref{EqBKT}. These correspond to the intersection between the stiffness parameter curves and the dashed orange line in the figure. Notably, we observe that the stiffness parameter is larger for parameters with larger superfluid gap, consistent with the behavior observed in Fig.~\eqref{Fig1}.}
\label{Fig2}
\end{figure}

Consistent with the results of Fig.~\eqref{Fig1}, we find in Fig.~\eqref{Fig2} that the stiffness parameter is larger for the same set of parameters for which the superfluid gap is larger. According to the BCS theory, we find that the stiffness and superfluid gap parameters are continuous functions of temperature. The stiffness parameter vanishes above a critical temperature $T_{\text{BCS}}$, indicated by the square marks in Fig.~\eqref{Fig2}.

{We should remark that our approach hinges on the concepts of majority and minority atoms; thus, this idea is no longer valid once the density of fermions is arbitrarily increased. In this case, the feedback of the Fermi gas to the BEC may change the properties of the collective excitations of the hard-core bosons.}

{\it Berezinskii–Kosterlitz–Thouless Superfluidity.-} It is well-established that two-dimensional systems cannot exhibit conventional long-range order. Instead, the transition is governed by the Berezinskii–Kosterlitz–Thouless theory (BKT). We obtain the BKT transition temperature from the expression:~\cite{berezinskii1972destruction,kosterlitz1973ordering,kosterlitz1974critical,Yin2014,Wu2016,camacho2016superfluidity} 
\begin{gather}
\label{EqBKT}
 J(T_{\text{BKT}})=\frac{2}{\pi}k_BT_{\text{BKT}}.
\end{gather}
Graphically, the BKT critical temperature is obtained from the crossing point of the orange dashed line, which represents $\frac{2}{\pi}k_BT$, and the stiffness parameter in Fig.~\eqref{Fig2}. From Fig.~\eqref{Fig2}, we observe that if the BCS critical temperature is small, then the BKT and BCS critical temperatures tend to agree (dot-dashed black line). However, as the BCS critical temperature increases, we observe significant deviations from the BKT theory prediction.

\begin{figure}[H]
\centering
\includegraphics[width=\columnwidth]{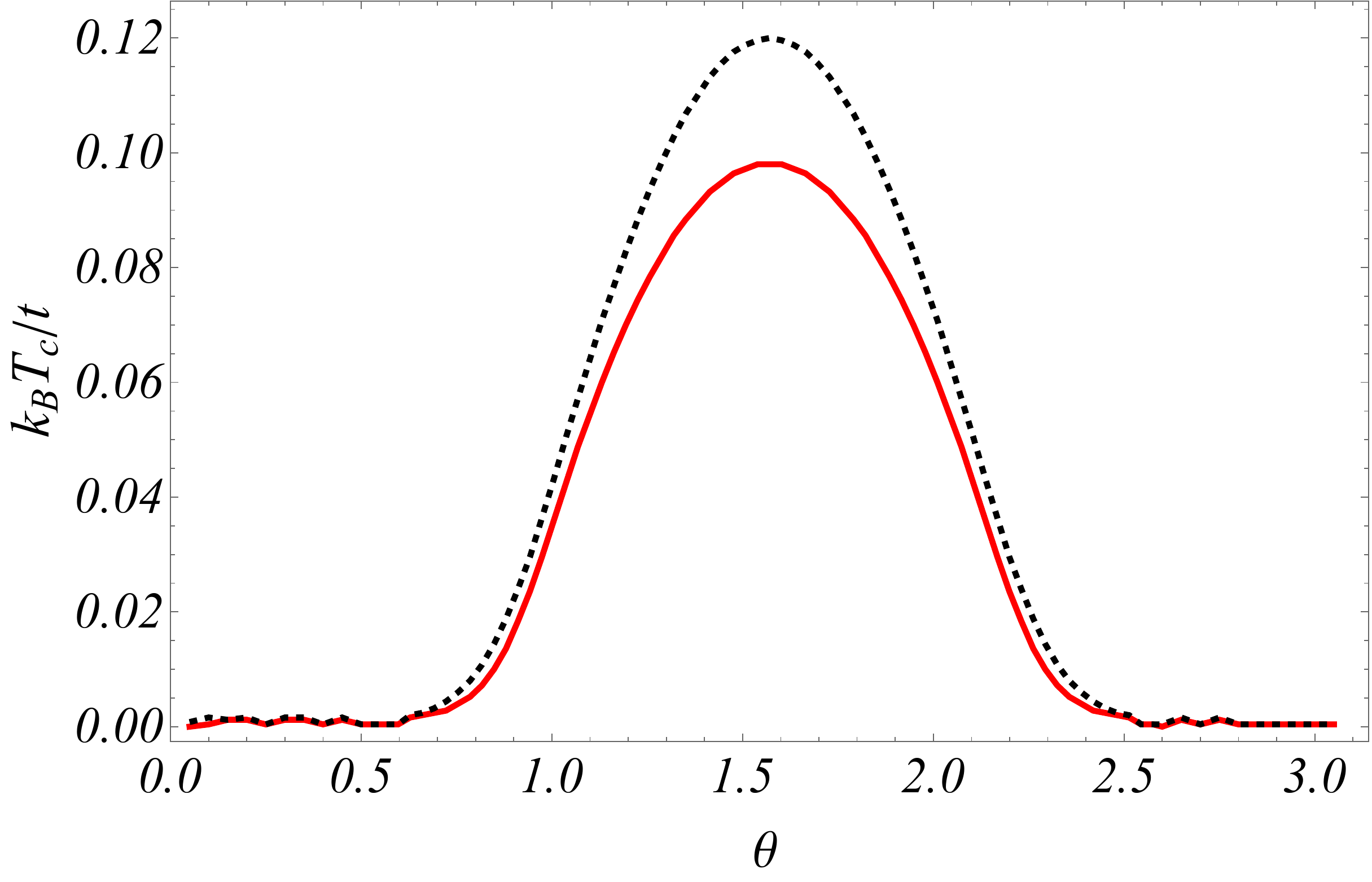}
\caption{Critical transition within the BKT formalism (solid line) and the BCS theory (dashed line) as a function of $\theta$ for $\mu/t=0$ (red). }
\label{Fig3}
\end{figure}

To quantitatively analyze the discrepancies between the BKT and BCS critical temperature, we plot in Fig.~\eqref{Fig3} the critical temperatures obtained from the BCS theory (dashed black) and the BKT formalism (solid red), as a function of $\theta$ for $\mu/t=0$ and $U_{ab}/t=2.5$. We observe that away from half-filling, the BCS critical temperature agrees remarkably well with the BKT prediction. However, as the BCS critical temperature increases closer to $\theta=\pi/2$, the BCS theory deviates from the BKT theory.

Figure~\eqref{Fig3} illustrates that the BKT critical temperature follows the same qualitative behavior as the superfluid gap in Fig.~\eqref{Fig2}. That is, the critical temperature is higher around half-filling of bosons than at small/large angles where the condensate fraction vanishes and the induced interaction becomes very inefficient. The maximal critical temperature is around $k_BT_{\text{BKT}}/t\approx 0.12$, which gives a typical temperature of several $nK$, which is of the order of magnitude of the state-of-the-art experiments with ultracold atoms.

\begin{figure}[h]
\centering
\includegraphics[width=\columnwidth]{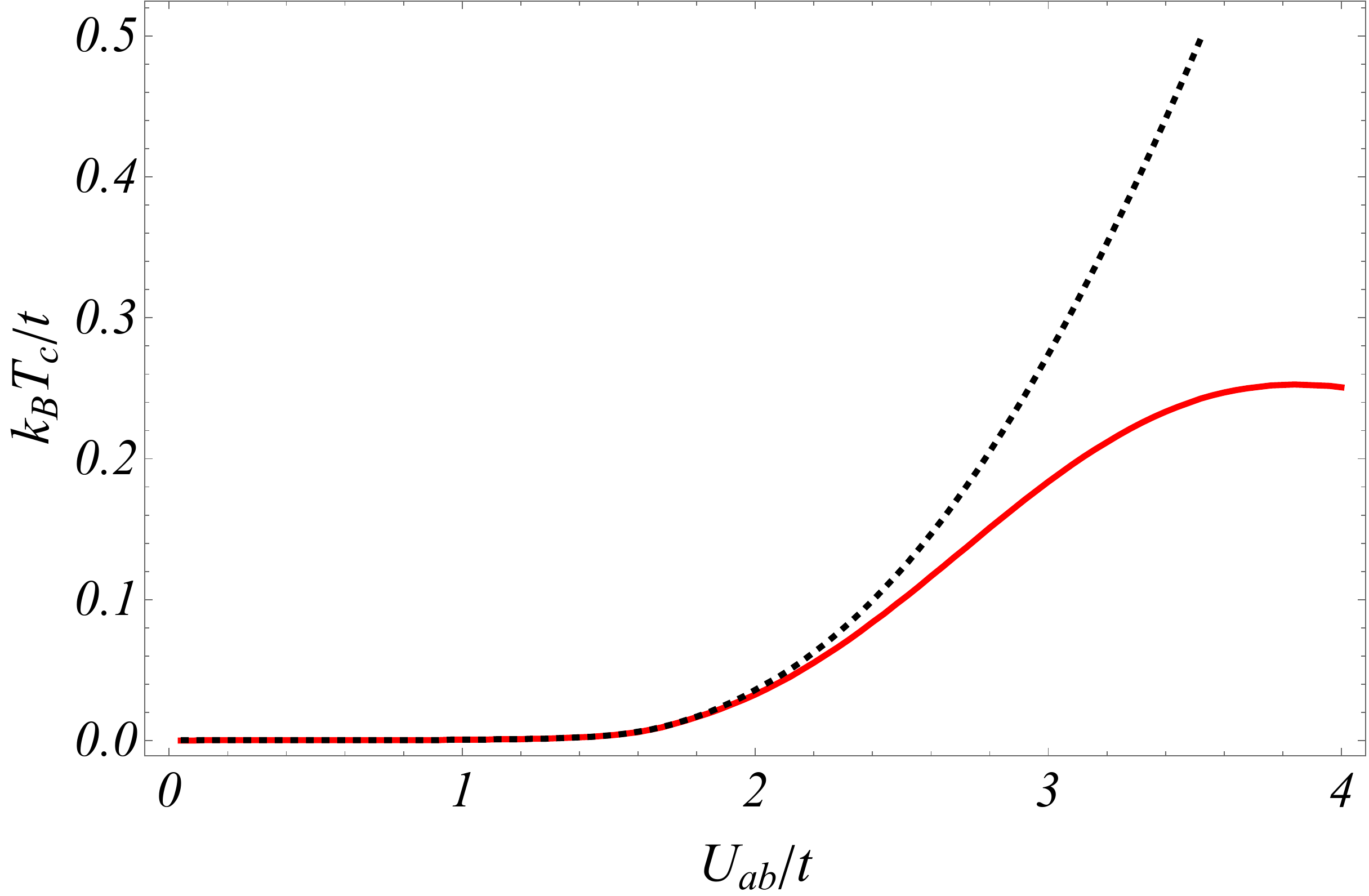}
\caption{Critical temperature as a function of $U_{ab}$. In black the BCS critical temperature while the red line gives the BKT critical temperature.}
\label{Fig4}
\end{figure}

Finally, we study the critical temperature as a function of the coupling strength $U_{ab}$ for fixed $\theta=\pi/2,$ and chemical potential $\mu/t=0.$ We plot in Fig.~\eqref{Fig4} the  BKT (solid red line) and the BCS  (dashed black line) critical temperature. We find that as long as $U_{ab}/t$ is kept small, both the BKT and BCS critical temperatures agree very well. With increasing $U_{ab}/t$, the critical temperature starts to deviate, while the BCS theory completely overestimates the transition temperature and increases without bound as $U_{ab}$ is increased, the BKT temperature saturates. {The BKT critical temperature, although rather modest, being of the order of magnitude of previous proposals~\cite{Midtgaard2017}, can potentially be enhanced by further increasing the fermion-boson interaction $U_{ab}$ . However, this requires to account for retardation effects and lies beyond the scope of the current manuscript. }

Conventional $s$-wave superfluidity can emerge as a consequence of magnon-like interactions which becomes maximal when the induced interaction is purely $s$-wave, that is, a mediated contact interaction. For weak boson-fermion interaction the BCS theory suffices to describe the critical temperature of the transition to superfluid. However, with increasing boson-fermion coupling, we obtain visible corrections to the BCS predictions via the BKT formalism.

{
\section{Experimental Considerations}
\label{Experiment}
Although our manuscript delves into the concrete field of quantum gases, our study is motivated and inspired by state-of-the-art experiments and the growing theoretical interest across three different fields: quantum gases, van der Waals heterostructures, and magnetic materials.

Quantum gases offer a powerful platform in which Bose-Hubbard models have been realized in low-dimensional geometries~\cite{Spielman2007,rubio2020floquet}. Moreover, the Fermi-Hubbard model has garnered significant experimental attention due to its potential role in high-temperature superconductors~\cite{Esslinger2010,mazurenko2017cold,mitra2018quantum,Hilker2023}. Several experiments have reported the ability to tune and control interactions and doping, crucial elements for the experimental realization of our theoretical proposal. Quantum microscopy provides exclusive opportunities for observation and manipulation, including the study of quantum states through local or nonlocal correlation functions.

Bose-Fermi-Hubbard models were first realized with a binary mixture of $^{87}$Rb and fermionic $^{40}$K~\cite{ospelkaus2006localization,Gunter2006,Best2009}. This mixture is particularly useful due to the existing Feshbach resonance, allowing for the tuning of fermion-boson interaction. Furthermore, the hard-core boson regime can be achieved by varying the hopping ratio of $^{87}$Rb atoms. Another potentially useful binary mixture is $^6$Li and $^{23}$Na, which also exhibit a Feshbach resonance~\cite{Schuster2012}. A comprehensive review of achieved quantum degenerate mixtures can be found in Ref.~\cite{schafer2020tools}, emphasizing the experimental feasibility of our proposal.

Van der Waals heterostructures have opened new avenues for realizing strongly correlated phases in periodically confined systems. In bilayers, moiré superlattices enable the realization of Hubbard models~\cite{Kennes2021, tang2020simulation}. Very recently, the first Bose-Fermi-Hubbard model was realized in these systems~\cite{gao2023excitonic}. Given the ability to tune Hubbard model parameters through the relative twist angle, we expect our study to draw attention towards realizing mediated superconductivity in these systems.

Finally, in magnetic materials, recent proposals suggest conventional and topological superconductivity mediated by skyrmions and other collective spin excitations~\cite{brataas2020spin,Rohling2018,johansen2019magnon}. This raises the potential for connections to quantum gases, where 'synthetic skyrmions' have already been produced~\cite{leslie2009creation,choi2012observation}. It remains an intriguing question whether their collective excitations can induce an interaction strong enough to support few-body states of a minority species.}

\section{Discussion and Conclusions}
\label{Conclusions}

Induced interactions play a crucial role in mediating collective phenomena in many quantum many-body systems. Magnon-induced interactions have emerged as a promising mechanism for generating exotic phases of matter. In this article, we demonstrate the analogue of spin wave excitations in an ultracold gas of hard-core bosons and show that the collective excitations are suitable for generating two-body bound states and mediating conventional superfluidity.

From a two-body perspective, we show that the filling factor of the bath can be used to control and manipulate the energy, number, and symmetry of the emerging two-body bound states. Although the bound states become more stable at half-filling, at this filling, only a single bound state emerges, and its symmetry is constrained to s-wave. Away from half-filling, we illustrate additional bound states with $s$, $p$, and $d$-wave symmetries.

Finally, we study mediated superfluidity within the BCS and BKT formalisms. We determine the gap parameter, superfluid stiffness, and BKT critical temperature and demonstrate that superfluidity becomes maximal close to the half-filling of bosons.

{Our study relies on several approximations we have discussed in the main text. First, the static approximation for the induced interaction. This assumes that the interaction is instantaneous and retardation effects can be neglected. That is, we expect  the speed of sound of the BEC to remain larger than the relevant velocities, that is, the collective modes are exchanged fast such that the mediated interaction can be regarded as static. 

The study of our proposal beyond the static approximation and for strong-coupling fermionic superfluidity stands as an interesting question to be addressed. In this context, our manuscript provides a valuable guide and a benchmark for these purposes.

}

Our study raises new questions related to the study of mediated interactions in van der Waals heterostructures, particularly moiré setups where Hubbard physics naturally appears, and the analysis of similar phases with Bose-Fermi mixtures of excitons and electrons~\cite{Kennes2021, JulkuA, gotting2022moire, yu2017moire, wu2018hubbard,Tan2020,PhysRevLettcam,tan2022bose,muir2022interactions,muir2022exciton,mistakidis2020many,julku2022light}. Another question relates to the study of mediated interactions when the underlying bath is in a strongly correlated phase such as a supersolid~\cite{JulkuA} and studies beyond the contact interaction for the boson-fermion direct interaction~\cite{Lahaye_2009,Chomaz_2023}.

{In homogeneous Bose-Fermi mixtures, induced interactions can lead to topological superconductivity when the Fermi atoms are spin-polarized. Thus an interesting avenue is to explore the nature of the superfluid for spin-polarized minority atoms.  }

\section{Acknowledgments} A. C. G. thanks R. Paredes for the reading of the manuscript.  M. S. G acknowledges to Consejo Nacional de Humanidades, Ciencias y Tecnolog\'ia (CONAHCYT) for the scholarship provided.  A.C.G. acknowledges financial support from UNAM DGAPA PAPIIT Grants No. IA101923 and  UNAM DGAPA PAPIME Grants No. PE101223.

\bibliography{ref}
\end{document}